# Path integral formulation of quantized fields

S. R. Vatsya
(York University, North York, Ontario, Canada)

## Abstract

Physical path integral formulation of motion of particles in Riemannian spaces is outlined and extended to deduce the corresponding field theoretical formulation. For the special case of a zero rest mass particle in Minkowski manifold, it is shown that the underlying space can be reduced to three-dimensional while similar formulation for a massive particle requires a four-dimensional structure. The solutions obtained by the present procedure are compared with the results obtained from its counterpart for a massive particle by setting the mass equal to zero. Although similar, the present formulation has some additional implications pertaining to the energy spectrum and the associated free field quantization, particularly it yields zero vacuum energy. The results are naturally extended to higher dimensions.

Subject Class: Quantum Physics

## I. Introduction

Path integral method is widely used to deduce the quantum mechanical analogues of the classical equations of motion resulting from an action principle as well as to study a variety of physical phenomena.[1,2] In particular, a generalized Schrödinger type equation with proper time as the evolution parameter has been deduced[3] for a particle, ignoring the case of half integral spin, coupled to an electromagnetic field by this method, which was first conjectured by Stückelberg.[4] Feynman also conjectured a similar equation and restricted its solutions to obtain the Klein-Gordon equation.[5] Solutions of the Stückelberg equation can be restricted in a parallel manner to derive the Klein-Gordon equation. Alternatively, a restriction on the allowed paths has been shown to restrict the admissible solutions yielding a set of equations, one of them being the Klein-Gordon equation.[3] Characterization of the admissible paths is naturally compatible with the properties of the trajectories in a Riemannian space parameterized by the arc-length.[6] These properties are used in the present article to obtain the Feynman type condition and thus, deriving the Klein-Gordon equation more precisely. A discrete mass spectrum results in the process.

The equation of motion for a particle of zero rest mass can be deduced from the Klein-Gordon equation by setting the mass equal to zero. However, some of the manipulations used in the path integral method are invalid for a zero rest mass particle. For a class of spaces, the motion of a zero rest mass particle can be formulated in a three-dimensional manifold with an appropriately selected evolution parameter, which eliminates the questionable manipulations. In the Minkowski manifold, this reduction replaces the proper time with time as the evolution parameter. The standard treatment then yields a Schrödinger type equation, which is reduced to a Klein-Gordon type equation with the rest mass equal to zero. The reduced equation is the same as the equation for the electromagnetic potentials resulting from Maxwell's equations in free space. However, the solutions obtained from the present formulation are periodic yielding a quantized energy spectrum with zero vacuum energy. In the process, the Minkowaskian space-time manifold is generated with the rest mass as the resulting invariant. The same procedure applies to a massive particle in the space-time manifold resulting in its quantized field formulation in a five-dimensional space. Mathematically, the results extend to higher dimensions, each time generating a new invariant parallel to the rest mass and a quantized field formulation, which may or may not describe any physical phenomena.

## II. Path integral method

In the framework of the path integral formulation of quantum mechanics, the wave function $\chi$ is defined as the sum of the phase factors over trajectories within a multiplicative constant, i.e.,

$$\chi(x, \tau(x)) = \sum_{z(\tau),\, y} K \exp[iS_{xy}(z(\tau); \tau(x), \tau(y))], \tag{1}$$

where $z(\tau) = \left(z^0(\tau),\, z^1(\tau),\, ..., z^{N-1}(\tau)\right)$ is a parameterization of the path in the underlying $N$-dimensional manifold. The analysis is valid for an arbitrary integer $N > 1$. The physical space-time corresponds to $N = 4$. The action $S_{xy}(z(\tau); \tau(x), \tau(y))$ is the classical action from $y$ to $x$ along an arbitrary path $z(\tau)$ obtained by integrating the Lagrangian from parameter value of $\tau(y)$ to $\tau(x)$ with $z(\tau(x)) = x$, $z(\tau(y)) = y$ and $K$ is a path-independent constant. The sum is over all curves originating at all points $y$ together with all parameter values. In this sense, the trajectories are in the $(N+1)$-dimensional manifold obtained by adjoining the parameter as the additional coordinate. The constant $K$ tends to infinite in the limit of all paths but the function $\chi(x, \tau(x))$ remains well defined. The composition property of $\chi$ is then used to obtain its integral representation.[1,2]

For a class of Lagrangians defined in terms of a suitable parameter $m$, usually the rest mass, the function $\hat{\chi}(x, \tau) = \chi(x, \tau) \exp(-im\tau/2]$ is more convenient for the computations involved. The properties of $\chi(x, \tau)$ and $\hat{\chi}(x, \tau)$ can be inferred from the properties of each other. An integral representation for $\hat{\chi}$ is given by[2,3,6]

$$\hat{\chi}(x, \tau + \varepsilon) = \int D(x) \left( \prod_{\mu=0}^{N-1} dy^\mu \right) \exp\left[iS(x, y, \varepsilon) - im\varepsilon/2\right] \hat{\chi}(y, \tau), \tag{2}$$

where $\tau = \tau(y)$, $\varepsilon = (\tau(x) - \tau(y))$. Explicit knowledge of the function $D(x)$, defined in terms of the constant $K$ and the properties of the manifold, is not required to derive the equation for the wave function,[6] which is obtained by expanding the left and right sides of Eq. (2) in powers of $\varepsilon$ and $(x-y)^\mu$ in the neighborhood of zero, respectively, and by comparing the first few terms. The present formulation, which is more suitable for the following analysis, differs slightly from Ref. [6]. The corresponding results can be inferred from each other or deduced following the same steps.

Since the difference between the actions along the classical and an arbitrary curve joining the two points is of the higher order, the action $S(x, y, \varepsilon)$ is defined with the required degree of accuracy to be the action from $y$ to $x$ along the classical trajectory, which can be obtained from the Hamilton-Jacobi equation. The procedure yields a Stückelberg type equation:

$$A \hat{\chi} = -2im \frac{\partial \hat{\chi}}{\partial \tau}, \qquad (3)$$

where the $\tau$-independent operator $A$ depends on the properties of the manifold and the Lagrangian. If the solution of Eq. (3) is restricted to the form

$$\hat{\chi}(x,\tau) = \hat{\psi}(x) \exp(-im\tau/2),$$

then the substitution yields the Klein-Gordon type equation

$$(A + m^2) \hat{\psi} = 0.$$

Feynman also conjectured an equation similar to Eq. (3) for a particle coupled to an electromagnetic field and adjusted the form of the solution accordingly to obtain the Klein-Gordon equation.[5]

In the above analysis, $\tau$ is an unrestricted parameter and the summation is over all trajectories. The Klein-Gordon equation is deduced by selecting appropriate solutions. Other solutions of the equations are ignored. The Klein-Gordon equation can be deduced more satisfactorily by restricting the admissible paths as follows. Let $S_{\hat{x}\hat{y}}(\rho)$ be the action associated with an arbitrary trajectory $\rho(\hat{x}\hat{y})$ joining the points $\hat{x}$ and $\hat{y}$. The admissible paths are defined by[3]

$$\kappa(\hat{x},\tau(\hat{x})) \exp\left[iS_{\hat{x}\hat{y}}(\rho)\right] \kappa^{-1}(\hat{y},\tau(\hat{y})) = 1, \qquad (4)$$

where $\kappa$ is a path-independent function. In a manifold without discontinuities, which will be assumed here, $\kappa = 1$. This characterization of the admissible trajectories has been incorporated in the path integral formulation as follows. If $x$, $y$ are two arbitrary points under consideration, then $z(\tau) = \rho$ is admissible in Eq. (1) if and only if $x$, $y$ are included in a segment $[\hat{x}, \hat{y}]$ of $z(\tau) = \rho(\hat{x},\hat{y})$ with $\hat{x}$, $\hat{y}$ and $\rho$ satisfying Eq. (4). Thus the set of contributing paths is characterized by $S_{\hat{x}\hat{y}}(\rho) = 2\pi n$, with $n$ being an arbitrary integer. Although not complete, this procedure to incorporate Eq. (4) into Eq. (1) is sufficiently accurate.[3,7] Also, some of the implications of Eq. (4) can be inferred directly.

Let $\chi'$ be defined by Eq. (1) with the set $\hat{S}(x,\tau(x))$ of contributing trajectories restricted by Eq. (4). The function $\hat{\chi}'(x,\tau) = \chi'(x,\tau) \exp[-im\tau/2]$ still admits the representation given by Eq. (2) and thus, satisfies Eq. (3). The set of admissible trajectories defined by Eq. (4) is dense in the set of all curves, eliminating the problem of unavailability of arbitrary, although small, variations of the coordinates required for this derivation that could arise otherwise. If the parameter leaves $\hat{S}(x,\tau(x))$ invariant under a translation $\tau \to (\tau + \hat{\tau})$, i.e., $\hat{S}(x,\tau(x)) = \hat{S}(x,\tau(x) + \hat{\tau})$, then from Eq. (1) $\chi'(x,\tau)$ is periodic with respect to $\tau$ with period $\hat{\tau}$, and the restriction on the admissible paths can be incorporated by selecting the periodic solutions of Eq. (3). This procedure has been used to derive a set of equations for the motion of a particle coupled to a gauge field[3] with $\hat{\tau} = 2\pi/m$, which results in the expansion

$$\chi'(x,\tau) = \sum_{n=-\infty}^{\infty} \psi_{n+1}(x) \exp[-imn\tau],$$

with $\hat{\chi}'$ satisfying Eq. (3), yielding the set of equations

$$[A + (2n+1)m^2] \psi_{n+1} = 0, \tag{5}$$

The case $n = 0$ corresponds to the Klein-Gordon equation.

Consider the case of the trajectories in a $N$-dimensional Riemannian space $\Re_N(\tau)$ with metric $g_{\mu\nu}$ and the Lagrangian $L = m (\dot{x}_\mu \dot{x}^\mu)^{1/2}$, where the dot denotes the derivative with respect to the arc-length $\tau$. An alternative Lagrangian $L' = [m (\dot{x}_\mu \dot{x}^\mu + 1)/2]$, together with $\tau$ as an independent parameter is more suitable for the applications of the methods of path integration and other pertaining analysis. The resulting equations can then be reduced by setting $\tau$ equal to the arc-length. Also, the results deduced here are independent of the sign of the Lagrangian, which will therefore be chosen according to convenience. In the physical space-time manifold, these Lagrangians correspond to the motion of a particle of rest mass $m$ with the arc-length $\tau$ as the evolution parameter.

The space $\Re_N(\tau)$ can be expressed as the union of two sub-manifolds $\Re_N^+(\tau)$ and $\Re_N^-(\tau)$ with a common boundary $\Re_N^0(\tau)$, characterized by $\tau^2 > 0$, $\tau^2 < 0$ and $\tau = 0$, respectively. Considerations will be restricted to $\Re_N^+(\tau)$. Parallel results apply to $\Re_N^-(\tau)$. In this case, Eq. (3) reduces to[6]

$$\left[\partial_\mu \partial^\mu + \frac{1}{3}R\right] \hat{\chi} = -2im \frac{\partial \hat{\chi}}{\partial \tau}, \tag{6}$$

where $R$ is the curvature scalar of the underlying manifold and $\tau$ is an independent parameter. Alternative arguments have been used in literature to obtain a different coefficient for the curvature term,[8,9] which is of no consequence for the present.

With $\tau$ as the arc-length, the action along all monotonic trajectories $\rho(xy)$ in $\Re_N^+(\tau)$ satisfies the equation

$$S_{xy}(\rho) = S(x, y, \tau(x), \tau(y)) = m (\tau(x) - \tau(y)). \tag{7}$$

Along non-monotonic trajectories the action can be expressed as the sum over its monotonic segments.[3] Eq. (7) defines a $N$-dimensional surface $\Im_N$ in a $(N+1)$-dimensional manifold $\Re_{N+1}(s)$ obtained by adjoining the arc-length to $\Re_N(\tau)$ as the additional coordinate $dx^N = d\tau$ with $m$ as its conjugate momentum. The corresponding metric $g'$ can be taken to be

$$g'_{NN} = -1, \; g'_{N\mu} = g'_{\mu N} = 0,; \; g'_{\mu\nu} = g_{\mu\nu}, \; \mu, \nu = 0, 1, ..., N-1.$$

The surface $\Im_N = \Re_{N+1}^0$ is defined by $s^2 = 0$, $\Re_{N+1}^+(s)$ by $s^2 > 0$ and $\Re_{N+1}^-(s)$ by $s^2 < 0$, where $s$ denotes the arc-length in $\Re_{N+1}(s)$ together with a new conjugate invariant $m'$, with same role in $\Re_{N+1}(s)$ as $m$ in $\Re_N(\tau)$. The union $\Re_{N+1}(s)$ of $\Re_{N+1}^+(s)$, $\Re_{N+1}^-(s)$ and their common boundary $\Re_{N+1}^0$, thus acquires a Riemannian structure with the same properties as $\Re_N(\tau)$. Thus, the above results admit natural extension to all dimensions.

Eq. (7) is just the statement of the fact that the infinitesimal arc-length $d\tau(y)$ is given by

$$dS(y) = S(y+dy, y) = m\sqrt{dy_\mu dy^\mu} = m\, d\tau(y). \qquad (8)$$

Eq. (7), equivalently Eq. (8), implies that that translation $\tau \to (\tau + 2\pi n/m)$ of the arc-length induces the translation $S \to (S + 2\pi n)$ of the action, where $n$ is an integer. Since Eq. (4) and thus, the set $\hat{S}(x, \tau(x))$, is invariant under the translation $S \to (S + 2\pi n)$, it is also invariant under the corresponding translation of the arc-length $\tau$. Hence, $\chi'(x, \tau)$ is periodic with period $\hat{\tau} = (2\pi/m)$, yielding the Riemannian space version of Eq. (5).[6] Since it will be of no consequence, the prime from $\chi'$ will be dropped. Also, it follows from Eqs. (1) and (7) or (8) that

$$\frac{\partial \chi(x,\tau)}{\partial \tau} = 0. \qquad (9)$$

Periodicity together with Eq. (9) determines $\chi$ precisely to be of the form:

$$\chi(x,\tau) = \psi_1(m;x), \text{ i.e., } \hat{\chi}(x,\tau) = \psi_1(m;x)\exp(-im\tau/2),$$

which together with Eq. (6) yields the Riemannian space version of the Klein-Gordon equation

$$\left[\partial_\mu \partial^\mu + m^2 + \frac{1}{3}R\right]\psi_1(m) = 0. \qquad (10)$$

While Eqs. (7) and (8) are satisfied in some other cases also,[3] this property is more transparent in a Riemannian space with the trajectories parameterized by the arc-length, making it naturally compatible with Eq. (4).

This procedure yields the Klein-Gordon equation for any given mass. However, the other harmonics associated with the periodicity are not redundant. In the original path-integral formulation, all parameter values are allowed. Eq. (4) together with the parameterization by the arc-length restricts the values to a discrete set corresponding to the harmonics, as follows. Let $S(m)$ denote the set of admissible trajectories in accordance with Eq. (4), with the associated period $[0, \hat{\tau}]$ with $\hat{\tau} = (2\pi/m)$. The set $S(m)$ is contained in $S(nm)$, which also contains the integral fractions of the trajectories in $S(m)$ and their unions. Thus, if a particle of mass $m$ is realizable, i.e., there are paths for it to follow, then its integral multiples are also realizable, classifying a discrete set of masses. The solutions $\psi_n(m) = \psi_1(nm)$ corresponding to the mass $(nm)$ are obtained from Eq. (10) by replacing $m$ by $(nm)$. The set of the solutions $\psi_n(m)$ can be represented collectively by

$$\psi(x,\tau) = \sum_{n=-\infty}^{\infty} \psi_n(m;x) \exp[inm\tau], \quad (11)$$

which satisfies the equation

$$\left[\partial_\mu \partial^\mu + \frac{1}{3}R\right]\psi = \frac{\partial^2 \psi}{\partial \tau^2}, \quad (12)$$

Conversely, Eq. (12), together with the boundary condition $\psi(x,0) = \psi(x,\hat{\tau})$ yields the set of the solutions $\{\psi_n(m)\}_{n=-\infty}^{\infty}$, establishing equivalence between the set of Klein-Gordon type equations, Eq. (10), and Eq. (12).

The set $S(m/n)$ contains $S(m)$ with the basic trajectory being a union of curves in $S(m)$ and the associated period equal to $(n\hat{\tau})$. Thus, the integral fractions of the basic mass are also realizable. However, they are classified in symmetry with the integral multiples in the conjugate Riiemannian space obtained by interchanging the roles of $x^\mu$ and their conjugate momenta. The conjugates of the momenta $p^\mu$ with respect to the conjugate space are $x_\mu$. The above steps can then be followed to yield $(n\hat{\tau})$, instead of $(nm)$, in complete symmetry.

An arbitrary trajectory in $\mathfrak{R}_N^+(\tau)$ has $N$ components implying that $\chi(x,\tau)$ as defined by Eq. (1) belongs to a $N$-dimensional space spanned by a set of functions. A basis can be generated for this space by restricting the trajectories to be summed over in Eq. (1). For example, let $Z^\nu(\tau) = z(\tau)$ with $z^\nu(\tau) = const.$ and let $\chi^\nu = \chi$ as defined by Eq. (1) with paths restricted to $Z^\nu(\tau)$. Then an arbitrary solution $\chi$ can be expressed as a linear combination of $\chi^\nu(x,\tau)$. Since the respective analysis is still valid, $\chi^\nu(x,\tau)$ satisfy Eq. (3) in general, and Eq. (6) in $\mathfrak{R}_N^+(\tau)$. While a basis is not unique, the function space spanned by any set is the same. Thus, the corresponding Klein-Gordon type equations, Eqs. (5) and (10), possess $N$-vector solutions. The dimension of the solutions of Eq. (12) is $(N+1)$, increase resulting from the added variable.

## III. Zero mass particle

Equation for a particle of zero rest mass can be formally obtained from Eq. (10) by setting $m = 0$. However, the arc-length in this case is identically equal to zero for all points and the derivation cannot be legitimately carried out following the above steps with mass and arc-length equal to zero. For a suitable underlying manifold, analysis parallel to the case of non-zero rest mass can be carried out for this case with the underlying manifold of dimension $(N-1)$, instead of $N$. While some of the following analysis is valid for more general Riemannian spaces, it will be restricted to a Minkowasian manifold, still denoted by $\mathfrak{R}_L(\square)$, where stronger results of significance can be obtained.

In a Minkowski space for a zero mass particle, the arc-length is zero and hence,

$$(dt)^2 = dx_\alpha dx^\alpha, \qquad \alpha = 1, 2, ..., N-1,$$

where $dt = dx^0 = dx_0$. This endowes the manifold $\Re_N^0$ with the structure of $\Re_{N-1}(t)$. The analysis of Sec. II can now be carried out in $\Re_{N-1}(t)$, which is Eucldian in this case, identifying the arc-length with time. The corresponding results can be deduced by replacing $\tau$ and $m$ with $t$ and $\omega$ respectively, in Sec. II, where $\omega$ is the conjugate momentum with respect to time, as $m$ is with respect to $\tau$. For physical space, $\omega$ coincides with the energy in the natural units presently being used. The counterpart of Eq. (12) reads

$$\partial_\mu \partial^\mu \phi = \left[ \frac{\partial^2}{\partial t^2} - \partial_\alpha \partial^\alpha \right] \phi = 0, \tag{13}$$

where the $N$- vector solution $\phi$ is periodic with $\hat{t} = (2\pi/\omega)$ as its period, i.e., $\phi(x, 0) = \phi(x, \hat{t})$. The solution $\phi$ thus, represents the solution for the set $\{n\omega\}_{n=-\infty}^{\infty}$ of the energies. The equation obtained by setting $m = 0$ in Eq. (10) for each $n$ is still Eq. (13) except for the periodic boundary conditions, and thus without the associated quantization. For distinction, the solutions without the periodicity will be denoted by $\phi'$.

Let $\xi = \phi, \phi'$ be a $N$-vector solution of Eq. (13). Define a tensor of rank two and its dual of rank $(N-2)$ by

$$f^{\mu\nu}(\xi) = \left( \xi^{\nu,\mu} - \xi^{\mu,\nu} \right), \quad \hat{f}^{\mu_1,...\mu_{N-2}}(\xi) = \delta^{\mu_1,...\mu_N} f_{\mu_{N-1},\mu_N},$$

respectively, where $\delta^{\mu_1,...\mu_N}$ is the Levi-Civita tensor density. By virtue of its definition, the dual $\hat{f}$ satisfies the Jacobi identities

$$\hat{f}^{\mu_1,...\mu_{N-2}}{}_{,\mu_j}(\xi) = 0, \quad j = 1, 2, ..., (N-2), \tag{14}$$

and $f^{\mu\nu}{}_{,\nu}(\xi) = \left( \xi^{\nu,\mu}{}_{,\nu} - \xi^{\mu,\nu}{}_{,\nu} \right)$, which together with the gauge fixing condition $\xi^\mu{}_{,\mu} = 0$ yields

$$f^{\mu\nu}{}_{,\nu}(\xi) = 0. \tag{15}$$

Conversely, Eq. (14) implies that $f$ can be expressed as the exterior derivative of a potential $\xi$ with $\hat{f}$ as its dual. Then Eq. (15) together with the gauge fixing condition yields Eq. (13) in $\xi$. Alternative gauge is sometime more convenient. The gauge used here corresponds to the Lorentz condition.

A more symmetric formulation results in the physical space where Eqs. (14) and (15) reduce to the Maxwell equations in free space. These equations are derived here by obtaining the equations for the potentials by the path integral method and then deducing the field equations. The standard procedure uses Maxwell's equations to obtain Eq. (13) in $\phi'$. The periodic boundary conditions deduced here induce the periodicity on $\phi$ and the associated quantization on $f^{\mu\nu}(\phi)$ and $\hat{f}^{\mu\nu}(\phi)$ also.

For the physical space, where $\phi$ is the electromagnetic potential vector, $(n\omega)$ is naturally interpreted as the energy of $n$ photons. For negative values of $n$, this corresponds to the absorption of $n$ photons and $n = 0$ represents the absence of photons, i.e., vacuum. Thus, in the present formulation, the vacuum energy is equal to zero.

In this section, a quantized field theory is formulated on $\mathfrak{R}_N^0$. Analogue of Eq. (7) and the corresponding construction then yields $\mathfrak{R}_N(\tau)$. The particle manifesting itself as zero mass particle in $\mathfrak{R}_N^0$ acquires a rest mass in $\mathfrak{R}_N(\tau)$. In Sec. II, a quantum mechanical particle theory corresponding to $\mathfrak{R}_N^+(\tau)$ is constructed in terms of the set of equations, Eq. (10), which is shown to be equivalent to Eq. (12) defined on $\mathfrak{R}_{N+1}^0$. By the method of the present section, a quantized field theory can be constructed starting with Eq. (12), on $\mathfrak{R}_{N+1}^0$, providing a field theoretical description of a particle in $\mathfrak{R}_N^+(\tau)$ as well as a new particle with the corresponding formulation in $\mathfrak{R}_{N+1}(s)$. The procedure induces the same structure on $\mathfrak{R}_{N+1}(s)$ as assumed in $\mathfrak{R}_N(\tau)$. For $N = 4$, this yields a theory of quantized free electromagnetic field on the surface of the light cone, and a particle theory in the space-time manifold for a particle of non-zero mass, which is defined in $\mathfrak{R}_4^+(\tau)$. This particle theory then yields the corresponding field theory in $\mathfrak{R}_5^0$. By the induction principle, the construction is valid for all $N \geq 4$. While mathematically a valid deduction, whether these would describe some physical phenomena, is not clear at present. Also, a field theory in terms of the tensors of rank two is symmetric with respect to the field and it's dual for $N = 4$. This symmetry cannot be preserved in any other dimension. Preservation of such symmetry may be possible for the fields described by the tensors of appropriate rank for even dimensions. Whether they would describe any physical fields or not is also not yet clear.

## IV. Harmonic oscillator

In this section, the present formulation is compared with the standard procedure to quantize the electromagnetic field. For this purpose, the following analysis will be restricted to the physical space where

$$\partial_\alpha \partial^\alpha = \nabla^2.$$

The equivalent gauge condition $\xi^0 = 0, \nabla \bullet \xi = 0$ is more convenient for this comparison. Eq. (13) with this gauge, is equivalent to the following equation for a harmonic oscillator

$$a_k''(\xi) + k^2(\xi) a_k(\xi) = 0, \tag{16}$$

where $a_k(\xi)$ are the Fourier coefficients of $\xi(t, x)$ with respect to the space coordinate $x$, and the prime denotes the derivative with respect to time $t$. The solutions of Eq. (16) satisfy the condition $k(\xi) \bullet a_k(\xi) = 0$ leaving two independent transverse waves. Both waves are essentially equivalent. Therefore $k$ will be assumed to represent one of the transverse waves. Periodicity of $a_k(\phi)$ with respect to $t$ with period $(2\pi/\omega)$ induces quantization for $k(\xi)$ according to $k_n(\phi) = n\omega$, in contradistinction with $a_k(\phi')$.

Standard quantization of the electromagnetic field proceeds with replacing these field oscillators with the quantum mechanical material oscillators with energies $(n+1/2)\omega$, implying a non-zero vacuum energy. Since the physical observations depend on the differences instead of the absolute values of the energy states, the vacuum energy is arbitrarily set equal to zero. Thus, the present formulation does not impact upon the corresponding results of quantum electrodynamics. However, the non-zero vacuum energy should result in gravitational effects, which are not observed.[10] Therefore, the present formulation results in a desirable improvement. However, this distinction between the field and the material oscillators requires further understanding.

For convenience, consider the classical equation for the non-relativistic one-dimensional harmonic oscillator and the associated Lagrangian,

$$q'' + \eta q = 0, \quad L_{nrel}(q',q) = \frac{1}{2}\left[q'^2 - \eta q^2\right], \tag{17}$$

respectively, where $q$ is the position of the oscillating particle of unit mass. The standard path integral method, which includes all trajectories as the contributing curves, yields the corresponding Schrödinger equation,

$$\frac{1}{2}\left[-\frac{\partial^2}{\partial q^2} + \eta q^2\right] u = i\frac{\partial u}{\partial t}. \tag{18}$$

The correspondence principle $p^\mu \to i\partial^\mu$ also yields the same equation. However, an implementation of the present restriction on the paths requires an appropriate parameter. It is shown below that the non-relativistic material oscillator is still adequately described by Eq. (18), distinguishing it from the field oscillators defined by Eq. (16).

Consider the corresponding relativistic harmonic oscillator represented by the Lagrangian

$$L_{rel}(\dot{t},\dot{q},q) = \frac{1}{2}\left[\dot{t}^2 - \dot{q}^2 + 1 + \eta q^2 \dot{t}\right]. \tag{19}$$

Although the action defined by this type of Lagrangians is not equal to the arc-length, the present method can still be implemented as follows.[3] In the rest frame of the set of the trajectories the action $S(x,y,\varepsilon) = \varepsilon$. In view of the Lorentz invariance of the system, the equality holds in the other frames as well, which implies that Eq. (8) is satisfied with the fundamental period $[0, 2\pi]$. This procedure is the same as used in the treatment of a relativistic fluid.[11] The pertaining results can now be obtained by the present method, which are available elsewhere also.[3] The results will be stated for a fixed mass $m=1$, for comparison. For this case the Klein-Gordon equation for the wave function $u$ reads

$$\left[\left(i\frac{\partial}{\partial t} - \frac{1}{2}\eta q^2\right)^2 + \frac{\partial^2}{\partial q^2}\right] u = u. \tag{20}$$

In the non-relativistic limit, Eq. (20) reduces to the Schrödinger equation, Eq. (18).

One-dimensional harmonic oscillator can also be formulated in a Riemannian space $\Re_3(\tau)$ following the Kaluza-Klein construction,[12] yielding[13]

$$\left[\left(i\frac{\partial}{\partial t} - \frac{1}{2}\eta q^2\right)^2 + \frac{\partial^2}{\partial q^2}\right] u = \left[1 - \frac{1}{3}R\right] u. \tag{21}$$

The curvature term $R = -f_{01}f^{01}/4 = (\eta q^2)^2/2$ in Eq. (21) results from the Riemannian structure of the Kaluza-Klein manifold, which in standard units is of higher order in Planck's constant than the other terms. Retaining the leading terms reduces Eq. (21) to Eq. (20), which reduces to Eq. (18) in the non-relativistic limit. Thus, a non-relativistic material oscillator in the space-time manifold is adequately described by the Schrödinger equation.

It is clear that the correspondence principle $p^\mu \to i\partial^\mu$ and the equivalent commutation relation $[x^\mu, p_\nu] = i\delta^\mu_\nu$ are preserved by the present formulation for the variables in a Riemannian manifold. Since the trajectories are adequately defined in the manifold spanned by $(t, a_k(\xi))$, the standard path integral method can still be used to deduce the Schrödinger equation, which is Eq. (18) with $a_k(\xi)$ replacing $q$, in accordance with the correspondence principle, yielding non-zero vacuum energy. The present formulation however, requires additional properties to preserve the correspondence. In the present article, the metrical structure of the underlying manifold is used to implement the restriction on the admissible trajectories. Alternative analysis may yield other methods to implement this restriction. In any case, since the present analysis yields zero vacuum energy, the correspondence in its original form cannot be preserved for the manifold spanned by $(t, a_k(\xi))$. Also, as the available techniques are not suitable for the present formulation, further extensions, e.g., the interaction of the radiation field and particles will require further developments.

## V. Concluding remarks

A procedure is developed in the present article to formulate a quantized field associated with a particle of zero mass on $\Re^0_N$ corresponding to a particle of non-zero mass in $\Re_{N-1}(\tau)$ for each $N \geq 4$. Mass here is defined as the conjugate to the arc-length in its respective space, acquiring usual meaning in the physical space-time. For $N = 4$, the particle in $\Re_3(\tau)$ is the physical particle moving about in space with its motion being tracked by the Euclidean arc-length instead of time. The corresponding field is the quantized electromagnetic field associated with photon in $\Re^0_4$, which is the surface of the light cone. The space $\Re_3(\tau)$ is naturally extended to $\Re_4(\tau)$ and the particle acquires a rest mass. Particle formulation of a particle in $\Re^+_4(\tau)$, i.e., a relativistic massive particle, is provided by the Klein-Gordon type equations and the corresponding field formulation in $\Re^0_5$. The procedure continues for all higher values of $N$.

For the present, a detailed analysis was restricted to the vector fields and particles in Minkowskian manifolds, although some of it is applicable to more general spaces. The concepts and techniques developed here can be used to study other particles and fields on the background of their respective spaces. Structures of some candidate manifolds are partly available in literature but some further developments may be required.

Starting with the observation of a particle in a region of space, the basic assumptions of the path-integral formulation supplemented with the restriction given by Eq. (4) are used here to deduce the structure of the space-time manifold and a quantized field formulation of the propagating electromagnetic field with zero vacuum energy. The Schrödinger and Klein-Gordon type equations are also deduced for this case, which have been derived elsewhere for physically more relevant cases [3-6]. Thus, the present formulation provides a unified framework for various components of the physical theories. The present formulation can be extended in various directions and it raises some interpretational issues, which are beyond the scope of the present article.